On the  attosecond laser pulse  tissue interaction

J. Marciak – Kozlowska

Institute of Electron Technology

Warsaw, Poland

M. Kozlowski

Physics Department Warsaw University

Warsaw, Poland

Abstract

The ultra-short laser pulses interaction with tissue is investigated. The hyperbolic diffusion thermal equation is developed and solved. It is shown that the femto and attosecond laser pulses generate the thermal wave with velocity of order of 10-3 m/s.

1. Introduction

Basically, laser pulse can interact with biological tissue in five ways: (a) the electromechanical mode, (b) ablation, (c) photothermal (congulative and vaporizing) processes, (d) photochemical (photodynamics) reaction, (e) biostimulation and wound heating.

In this paper we will study in detail the photothermal processes. In the first approximation the tissue is the semiconductor and the electromagnetic field of the laser pulse generates the transport of charge, mass and heat in rissue. These transport phenomena in first approximation can be described as the diffusion processes [1].

It is well known that the application of the Fourier law to the analysis of the diffusion is limited to the case when the pulse duration is much longer than the relaxation time for the transport processes. The contemporary femtosecond and attosecond lasers open the new possibility for the study of the laser – tissue interaction. As the both duration of the pulses: 1 fs and 1 as are much shorter than the relaxation times the Fourier law can not be valid.

As was shown in monograph [2] for $\Delta t$ (pulse duration) $\ll \tau$ (relaxation time) the master equation for the temperature field is the hyperbolic diffusion equation, Heaviside equation:

$$\tau \frac{\partial^2 T}{\partial t^2} + \frac{\partial T}{\partial t} = D_T \Delta T, \tag{1}$$

where $T(\vec{r},t)$ is the temperature field, $\tau$ is the relaxation time and $D_T$ is the thermal diffusion coefficient.

For the hypebolic diffusion in the field of the potential $V(\vec{r})$ the hyperbolic diffusion has the form [2]:

$$\tau\frac{\partial^2 T}{\partial t^2}+\frac{\partial T}{\partial t}+\frac{2V}{Dm}T=D\nabla^2 T. \tag{2}$$

In the general case the Eq.(2) is the hyperbolic non – linear partial equation which describes the transport of heat in laser – pulse – tissue interaction. The Eq.(2) for $\tau=0$ Describes the diffusion processes. On the other hand for $\tau\to\infty$ Eq.(2) is the master equation for the thermal wave:

$$\tau\frac{\partial^2 T}{\partial t^2}+\frac{2V}{Dm}T=D\nabla^2 T(\vec{r},t). \tag{3}$$

From above we conclude that the Eq.(2) presents the *unified* description of the thermal processes generated in tissue: diffusion and wave model.

In the case of the source term $\Phi_1(\vec{r},t)$ equation (2) has the form

$$\tau\frac{\partial^2 T}{\partial t^2}+\frac{\partial T}{\partial t}+\frac{2V}{Dm}T=D\nabla^2 T+\Phi_1(\vec{r},t). \tag{4}$$

or

$$\frac{\partial^2 T}{\partial t^2}+\frac{1}{\tau}\frac{\partial T}{\partial t}=\frac{D}{\tau}\nabla^2 T-\frac{2V}{Dm\tau}T+\frac{\Phi_1(\vec{r},t)}{\tau}. \tag{5}$$

Let us introduce the abrevation

$$\begin{aligned}\frac{1}{\tau}&=k;\\ \frac{D}{\tau}&=a^2;\\ -\frac{2V}{Dm\tau}&=b;\\ \frac{\Phi_1(\vec{r},t)}{\tau}&=\Phi(\vec{r},t).\end{aligned} \tag{6}$$

Considering formulae (6) Eq.(5) can be written as

$$\frac{\partial^2 T}{\partial t^2}+k\frac{\partial T}{\partial t}=a^2\nabla^2 T+bT(\vec{r},t)+\Phi(\vec{r},t). \tag{7}$$

2. Solution of the one dimentional hyperbolic diffusion with source term $\Phi(x,t)$

In the following we will obtain the solution of Eq.(7) for constant $D$ and $\tau$. To that aim let us consider the Cauchy problem:

Domain: $-\infty < x < \infty$.

Initial conditions are prescribed:

$$T(x,0) = f(x)$$
$$\frac{\partial T}{\partial x}(x,0) = g(x). \tag{8}$$

Solution for $b + \frac{1}{4}k^2 = c^2 > 0$ [3]:

$$T(x,t) = \frac{1}{2}\exp\left(-\frac{1}{2}kt\right)\int_{x-at}^{x+at} \frac{I_1\left(c\sqrt{t^2-(x-\varsigma)^2/a^2}\right)}{\sqrt{t^2-(x-\varsigma)^2/a^2}} f(\varsigma)d\varsigma$$
$$+ \frac{1}{2a}\exp\left(-\frac{1}{2}kt\right)\int_{x-at}^{x+at} I_0\left(c\sqrt{t^2-(x-\varsigma)^2/a^2}\right)\left[g(\varsigma)+\frac{1}{2}kf(\varsigma)\right]d\varsigma$$
$$+ \frac{1}{2a}\int_0^t \int_{x-v(t-\eta)}^{x+a(t+\eta)} \exp\left(-\frac{1}{2}k(t-\eta)\right)I_0\left(c\sqrt{(t-\eta)^2-(x-\varsigma)^2/a^2}\right)\Phi(\varsigma,\eta)d\varsigma d\eta \tag{9}$$
$$+ \frac{1}{2}\exp\left(-\frac{1}{2}kt\right)[f(x-at)+f(x+at)]$$

Solution for $b + \frac{1}{4}k^2 = -c^2 < 0$ [3]:

$$T(x,t) = \frac{1}{2}\exp\left(-\frac{1}{2}kt\right)[f(x-at)+f(x+at)]$$
$$- \frac{ct}{2a}\exp\left(-\frac{1}{2}kt\right)\int_{x-at}^{x+at} \frac{J_1\left(c\sqrt{t^2-(x-\varsigma)^2/a^2}\right)}{\sqrt{t^2-(x-\varsigma)^2/a^2}} f(\varsigma)d\varsigma$$
$$+ \frac{1}{2a}\exp\left(-\frac{1}{2}kt\right)\int_{x-at}^{x+at} J_0\left(c\sqrt{t^2-(x-\varsigma)^2/a^2}\right)\left[g(\varsigma)+\frac{1}{2}kf(\varsigma)\right]d\varsigma$$
$$+ \frac{1}{2a}\int_0^t \int_{x-a(t-\eta)}^{x+a(t+\eta)} \exp\left(-\frac{1}{2}k(t-\eta)\right)J_0\left(c\sqrt{(t-\eta)^2-(x-\varsigma)^2/a^2}\right)\Phi(\varsigma,\eta)d\varsigma d\eta.$$

(10)

Formulae (9) and (10) describe the propagation of the thermal disturbance $f(x \pm vt)$ with velocity

$$v = a = \sqrt{D/\tau}. \tag{11}$$

Considering that diffusion coefficient *D* for haemoglobin in water is of the order of $10^{-11}$ m²/s and relaxation time ~ 170 *μ*s [3] the velocity of the thermal wave for haemoglobin is of the order

$$v \sim 10^{-3} \text{ m/s} \tag{12}$$

From formulae (10 – 12) we conclude that in tissue ultra-short laser pulse generates the thermal wave with velocity of the order $10^{-3}$ m/s. The mean free path for the thermal wave in haemoglobin is of the order of

$$\lambda = v \cdot \tau \approx 10^{-7} m \sim 10^{-1} \mu m.$$

From medical points of wiev the application of ultra-short laser pulses offers the very good localization of the thermal energy. In tissue radiated by ultra-short laser pulses the thermal "hot spot" with radius of the order of 0.1 *μ*m can be generated.

Conclusions

In this paper the new equation for thermal phenomena in tissue is proposed and solved. It was shown that for laser pulse duration < relaxation time in tissue the thermal wave can be generated with speed of the order of $10^{-3}$ m/s.